\documentclass{PoS-hep}
\usepackage{amsmath}
\usepackage{bm}
\usepackage{epsfig}
\usepackage{graphics}
\usepackage{upgreek}

\renewenvironment{subequations}{%
\refstepcounter{equation}%
\setcounter{parentequation}{\value{equation}}%
  \setcounter{equation}{0}
  \ignorespaces
}{%
  \setcounter{equation}{\value{parentequation}}%
  \ignorespacesafterend
}

\newcommand{\beqs}{\begin{subequations}}
\newcommand{\eeqs}{\end{subequations}}
\newcommand{\eec}{\end{center}}
\newcommand{\bec}{\begin{center}}
\newcommand{\eem}{\end{matrix}}
\newcommand{\bem}{\begin{matrix}}
\newcommand{\Eref}[1]{Eq.~(\ref{#1})}
\newcommand{\Sref}[1]{Sec.~\ref{#1}}
\newcommand{\Fref}[1]{Fig.~\ref{#1}}
\newcommand{\Tref}[1]{Table~\ref{#1}}
\newcommand{\cref}[1]{Ref.~\cite{#1}}

\newcommand\eqs[2]{Eqs.~(\ref{#1}) and (\ref{#2})}

\newcommand{\sEref}[2]{Eq.~(\ref{#1}{\ftn\sf {#2}})}

\newcommand{\eeq}{\end{equation}}
\newcommand{\beq}{\begin{equation}}
\newcommand{\ba}{\begin{array}}
\newcommand{\ea}{\end{array}}
\newcommand{\bea}{\begin{eqnarray}}
\newcommand{\eea}{\end{eqnarray}}
\newcommand{\ftn}{\footnotesize}

\newcommand{\ssz}{\scriptsize}

\newcommand{\etal}{{\it et al.\/}}
\newcommand{\tr}{{\mbox{\sf\ssz T}}}
\renewcommand{\Im}{{\mbox{\sf\ftn Im}}}
\renewcommand{\Re}{{\mbox{\sf\ftn Re}}}

\newcommand{\hepph}[1]{{\ftn \tt hep-ph/#1}}

\newcommand{\astroph}[1]{{\ftn \tt astro-ph/#1}}
\newcommand{\arxiv}[1]{{\ftn\tt  arXiv:#1}}
\newcommand{\ns}{\ensuremath{n_{\rm s}}}

\newcommand\vev[1]{\langle {#1} \rangle}
\def\lf{\left(}
\def\rg{\right)}

\newcommand{\Gr}{\ensuremath{\widetilde{G}}}
\newcommand{\Trha}{\ensuremath{T_{\rm 1rh}}}
\newcommand{\Trhb}{\ensuremath{T_{\rm 2rh}}}
\newcommand{\Vhi}{\ensuremath{V_{\rm HI}}}
\newcommand{\Vhio}{\ensuremath{V_{\rm HI0}}}
\newcommand{\Vpq}{\ensuremath{V_{\rm PQ}}}
\newcommand{\Vpqo}{\ensuremath{V_{\rm PQ0}}}
\newcommand{\ck}{{\ensuremath\mbox{\sl a}}}
\newcommand{\cks}{{\mbox{\ftn\sl a}}}

\newcommand{\mP}{\ensuremath{m_{\rm P}}}
\newcommand{\ld}{\ensuremath{\lambda}}
\newcommand{\ka}{\ensuremath{\kappa_a}}
\newcommand{\kp}{\ensuremath{\kappa}}
\newcommand{\GeV}{{\mbox{\rm GeV}}}

\newcommand{\snH}{\ensuremath{\nu^c_\Phi}}
\newcommand{\snHb}{\ensuremath{\bar\nu^c_\Phi}}
\newcommand{\seH}{\ensuremath{e^c_\Phi}}
\newcommand{\seHb}{\ensuremath{\bar e^c_\Phi}}

\newcommand{\sg}{\ensuremath{\sigma}}

\newcommand{\msp}{\ensuremath{m^2_{+}}}
\newcommand{\msm}{\ensuremath{m^2_{-}}}
\newcommand{\mspm}{\ensuremath{m^2_{\pm}}}

\title{Combining F-Term Hybrid Inflation With a Peccei-Quinn Phase Transition }

\ShortTitle{Combining FHI with a PQ Phase Transition }

\author{\speaker{C. Pallis}\\
        Department of Physics, University of Cyprus, \\
        P.O. Box 20537, Nicosia 1678, CYPRUS\\
        E-mail: \email{cpallis@ucy.ac.cy}}

\abstract{We consider an inflationary model based only on
renormalizable superpotential terms in which a superheavy scale
\emph{F-term hybrid inflation} (FHI) is followed by a
\emph{Peccei-Quinn} (PQ) phase transition. We show that the field
which triggers the PQ phase transition influences drastically the
inflationary dynamics and that the Universe undergoes a secondary
phase of reheating after the PQ phase transition. Confronting FHI
with the current observational data we find that, for the central
value of the spectral index, the grand unification scale can
assume its supersymmetric value for more or less natural values
for the remaining model parameters. On the other hand, the final
reheat temperature after the PQ phase transition turns out to be
low enough to avoid the gravitino problem.  \\ \\ {\sl\bfseries
Published in}~~{PoS (CORFU2011) 028}.}

\FullConference{Proceedings of the Corfu Summer Institute 2011\\
 "School and Workshops on Elementary Particle Physics and Gravity"\\
 September 4-18, 2011\\ Corfu, Greece}

\begin{document}

\section{Introduction}

In this talk, which is based on \cref{pqhi}, we describe how we
can achieve a cosmological scenario in which a superheavy
\emph{F-term hybrid inflation} (FHI) is followed by a
\emph{Peccei-Quinn phase transition} (PQPT) using two similar
renormalizable superpotential terms. Below, we first briefly
review the basic ingredients of our construction in Sec.~\ref{Fhi}
and \Sref{pqpt} and outline the structure of our proposal in
\Sref{plan}.

\subsection{F-term Hybrid Inflation}\label{Fhi}

One of the most natural, popular and well-motivated inflationary
model is the \emph{supersymmetric} (SUSY) FHI \cite{hybrid,
susyhybrid,leptohybrid}. It can be realized adopting the
superpotential
\begin{equation}
\label{Whi} W_{\rm FHI}=\kappa S\left(\bar \Phi\Phi-M^2\right),
\end{equation}
which is consistent with a continuous $R$-symmetry
\cite{susyhybrid} under which
\begin{equation}\label{Rsym} S\  \to\ e^{i\alpha}\,S,~\bar\Phi\Phi\ \to\
\bar\Phi\Phi,~W_{\rm FHI} \to\ e^{i\alpha}\, W_{\rm FHI}.
\end{equation}
Here, $S$ is a \emph{left-handed} (LH) superfield, singlet under a
\emph{grand unified theory} (GUT) gauge group $G$; $\bar{\Phi}$
and $\Phi$ is a  pair of LH superfields belonging to non-trivial
conjugate representations of $G$, and reducing its rank by their
\emph{vacuum expectation values} (v.e.vs); $\kappa$ and $M$ are
parameters which can be made positive by field redefinitions.

The SUSY potential induced by $W_{\rm FHI}$ in Eq.~(\ref{Whi})
along the D-flat direction $\vert\bar{\Phi} \vert=\vert\Phi\vert$
is
\beq V_{\rm FHI}=\kappa^2\left|\bar\Phi\Phi-M^2\right|^2+
\kappa^2|S|^2\lf|\bar\Phi|^2+|\Phi|^2\rg.\label{Vfhi}\eeq
$W_{\rm FHI}$ gives rise to FHI, since there is a F-flat
direction, with $\bar{\Phi}=\Phi=0$ and constant potential energy
$V_{\rm F}\simeq\kappa^2M^4$, which is a local minimum of $V_{\rm
F}$ for $S>M$. Also, $W_{\rm FHI}$ leads to the spontaneous
breaking of $G$, since the SUSY vacuum lies at
\beq\langle S\rangle=0~~\mbox{and}~~\vert\langle\bar{\Phi}
\rangle\vert=\vert\langle\Phi\rangle\vert= M,\label{fhivev}\eeq
with the non-zero v.e.vs of $\bar{\Phi}$ and $\Phi$ developed
along the \emph{Standard Model} (SM) singlet directions.

One of the shortcomings of FHI is the tension which, in general,
exists between the predicted (scalar) spectral index $n_{\rm s}$
and the recent seven-year results \cite{wmap} from the
\emph{Wilkinson microwave anisotropy probe} (WMAP7) satellite.
Indeed, it is well-known that the realization of FHI within
minimal \emph{Supergravity} (SUGRA) leads to $\ns$ which is just
marginally consistent with the fitting of the WMAP7 data by the
standard power-law cosmological model with \emph{cold dark matter
and a cosmological constant} ($\Lambda$CDM). One possible
resolution of this problem is \cite{gpp} the addition to the
K\"ahler potential of a non-minimal quatric term of the inflaton
field with a convenient choice of its sign. As a consequence, a
negative mass term for the inflaton is generated. In the largest
part of the parameter space, the inflationary potential acquires a
local maximum and minimum. Then, FHI of the hilltop type
\cite{lofti} can occur as the inflaton rolls from this maximum
down to smaller values. Therefore, $\ns$ can become consistent
with data, but only at the cost of an extra indispensable mild
tuning \cite{gpp} of the initial conditions. Another possible
complication is that the system may get trapped near the minimum
of the inflationary potential and, consequently, no FHI takes
place.

\subsection{Supersymmetrizing the PQ Solution to the Strong CP
Problem}\label{pqpt}

Due to the non-perturbative structure of the vacuum of $
SU(3)_{\rm C}$ the lagrangian of \emph{quantum chromodynamics}
(QCD) includes a CP-violating term, involving the strong coupling
constant, $g_3$, the gluon field-strength tensor, ${\mathcal G}$,
and its dual, $\widetilde {\mathcal G}$. I.e.,
\beq {\cal L}_{\rm QCD}={g_3^2\over32\pi^2}\bar\theta\ {\mathcal
G}^{\rm a\mu\nu}\ \widetilde {\mathcal G}^{\rm
a}_{\mu\nu}+\cdots~~\mbox{with}~~\bar\theta\lesssim5
\cdot10^{-10},\label{Lcp}\eeq
since $\bar\theta$ is involved in the computation of the neutron
electric dipole moment which is experimentally determined, with
result
\beq d_{\rm n}\simeq4.5\cdot10^{-16}\ \bar\theta <2.9\cdot10^{-26}
~\mbox{e-cm~~~at~~90\% c.l.}\label{dn}\eeq
The smallness of $\bar\theta$ consists the infamous strong CP
problem. The most promising solution, proposed \cite{pq} by Peccei
and Quinn, is to introduce a global color anomalous $U(1)_{\rm
PQ}$ symmetry which is spontaneously broken at an energy scale
$f_a \simeq (10^{9}-10^{12})~\GeV$, known as PQ energy scale. The
Goldstone boson, $a(x)$, associated with such symmetry breaking is
called axion. The Lagrangian term resulting after the spontaneous
symmetry breaking of the $U(1)_{\rm PQ}$ symmetry reads:
\beq {\cal L}_{a}={1\over2}\partial^\mu a\partial_\mu a+
c_a{g_3^2\over32\pi^2}{a\over f_a}{\mathcal G}^{\rm
a}_{\mu\nu}\widetilde {\mathcal G}_{\rm a\mu\nu},\label{La}\eeq
where $c_a$ is a model-dependent parameter. When considering the
total lagrangian parts of \eqs{Lcp}{La}, an effective potential
for $a$ appears, whose minimum is reached when the so-called
(axion) misalignment angle vanishes, i.e.,
\beq \theta=\bar\theta+c_a{a\over f_a}=0
~~\mbox{or}~~<a>=-\bar\theta {f_a\over c_a}\>\cdot\label{veva}\eeq
Therefore, minimizing the potential with respect to $a$ sets the
offending CP-violating term to zero. Essentially, $\bar \theta$ is
promoted to a dynamical variable that evolves to its CP-conserving
minimum, $\theta=0$, where $\theta$ can be seen as the phase of a
new complex scalar field, named PQ field.

Within a SUSY framework, the spontaneous breaking of $U(1)_{\rm
PQ}$ can be obviously realized adopting \cite{goto} a
renormalizable superpotential, $W_{\rm PQ}$, similar to that of
\Eref{Whi} where $S$ is replaced by another $G$ and PQ singlet LH
superfield, $P$, with the same $R$ charge, while $\Phi$ and $\bar
\Phi$ are replaced by a pair of $G$ singlet oppositely PQ-charged
LH superfields, $\bar Q$ and $Q$. Indeed, the superpotential
\beq W_{\rm PQ}= \kappa_a P\left(\bar QQ
-f_a^2/4\right),\label{Wpq}\eeq
is invariant under the $U(1)_{\rm PQ}$ transformations
\beq P\ \to\ P,\>Q\ \to\ e^{i\alpha}\,Q, \bar Q\ \to\
e^{-i\alpha}\,Q\eeq
and lead to the F-term SUSY potential
\beq V_{\rm PQF}=\kappa_a^2\left|\bar Q Q-f^2_a/4\right|^2\ +\
\kappa^2_a\left|P\right|^2\left(|\bar Q|^2+
|Q|^2\right),\label{Vpqf}\eeq
from where we can infer that $U(1)_{\rm PQ}$ can be spontaneously
broken due to the following v.e.vs:
\beq
\vev{P}\simeq0~~\mbox{and}~~\vev{\phi_Q}=f_a~~~\mbox{with}~~~2Q=2{\bar
Q}=\phi_Q\label{PQvev}\eeq
-- since the sum of the arguments of $\vev{\bar Q}$ and $\vev{Q}$
must be $0$, $\bar Q$ and $Q$ can be brought to the real axis by
an appropriate PQ transformation. In reality, however, the total
potential of the PQ fields is
\beq V_{{\rm PQ}a}=V_{\rm
PQF}+V_a,~~~\mbox{where}~~~V_a=m^2_a(T)f_a^2(1-\cos\theta)\label{Vapq}\eeq
(with $m_a$ the temperature, $T$, dependent $a$ mass) comes from
nonperturbative QCD effects associated with instantons
\cite{sikivie}, that break explicitly $U(1)_{\rm PQ}$ down to a
$\mathbb{Z}_N$ discrete subgroup, where $N$ is the sum of the PQ
charges of the $SU(3)_{\rm C}$ triplets and antitriplets of the
model. Therefore, the breakdown of $\mathbb{Z}_N$ by the v.e.vs in
\Eref{PQvev} may lead \cite{sikivie} to cosmologically
catastrophic domain walls which, however, can be avoided
\cite{georgi} by introducing extra matter superfields -- see
\Sref{walls}.

A by-product of the $U(1)_{\rm PQ}$ spontaneous breaking is that
we can achieve \cite{PQmu} a resolution of the $\mu$-problem of
MSSM by considering, e.g., a non-renormalizable superpotential
term of the form $\bar Q^2H_uH_d/\mP$, which after the spontaneous
breakdown of $U(1)_{\rm PQ}$ leads to the $\mu$ term of the MSSM,
with $|\mu|\sim\lambda_\mu\left|\vev{\bar Q}\right|^2/m_{\rm P}$,
which is of the right magnitude if $\left|\vev{\bar
Q}\right|=f_a/2\simeq 5\cdot 10^{11}~{\rm GeV}$ and
$\lambda_\mu\simeq(0.001-0.01)$ -- here, $m_{\rm P}\simeq
2.44\cdot10^{18}~{\rm GeV}$ is the reduced Planck scale; $H_u$ and
$H_d$ are the electroweak Higgses of MSSM which couple to up- and
down-type quarks respectively.

\subsection{Outline}\label{plan}

The key point of our attempt in combining both ingredients (FHI
and PQPT) described in Sec.~\ref{Fhi} and \ref{pqpt} is that $P$
can be regarded as the linear combination of the $G\times
U(1)_{\rm PQ}$ singlets with the $R$ charge of the superpotential
that does not couple to $\bar\Phi\Phi$ -- cf.
Ref.~\cite{tetradis}. As a consequence, an unavoidable
superpotential coupling $S\bar Q Q$ and a $S-P$ mixing in the
K\"ahler potential arise -- see Sec.~\ref{fhim}. These facts
influence drastically the inflationary set-up described in
\Sref{fhi}. In addition, the value of $P$ after FHI is to be kept
larger than $f_a/2$ so as to achieve an instantaneous domination
of the PQ system over radiation in order to alleviate the
gravitino ($\Gr$) problem \cite{brand, kohri}. These effects are
presented in \Sref{pfhi}. We end up testing our model against
observations in Sec.~\ref{cont} and summarizing our results in
Sec.~\ref{con}.

\section{Model Description}\label{fhim}

We below describe the structure of our model in \Sref{m1}, we
sketch its cosmological consequences in \Sref{m2} and explain how
we avoid the formation of domain walls in Sec.~\ref{walls}.

\subsection{The General Set-up}\label{m1}

In order to explore our scenario, we identify $G$ with the
left-right symmetric gauge group $G_{\rm LR} = SU(3)_{\rm C}\times
SU(2)_{\rm L}\times SU(2)_{\rm R} \times U(1)_{B-L}$, which can be
broken down to the SM gauge group $G_{\rm SM}=SU(3)_{\rm C}\times
SU(2)_{\rm L} \times U(1)_{Y}$ through the v.e.vs acquired by a
conjugate pair of $SU(2)_{\rm R}$ doublet Higgs, $\bar\Phi$ and
$\Phi$. As a consequence, no cosmic strings are produced in the
end of FHI and, therefore, no extra restrictions on the parameters
have to be imposed -- c.f. Ref.~\cite{mairi}. The model possesses
also three global $U(1)$ symmetries. Namely, a (color) anomalous
$R$ symmetry ${U(1)_R}$, an anomalous PQ symmetry ${U(1)}_{\rm
PQ}$ and the baryon number symmetry ${U(1)}_B$. The
representations under $G_{\rm LR}$ and the charges under the
global symmetries of the various matter and Higgs superfields are
presented in Table~\ref{tabLR}, which also contains $n$ extra
matter superfields ($\bar D_{\rm a}-D_{\rm a}$ and $H_{\rm a}$)
required for evading the domain-wall problem associated with PQPT
together with a new imposed global ${U(1)}_D$ symmetry  -- see
Sec.~\ref{walls}.

\renewcommand{\arraystretch}{1.2}
\begin{table}[!t]
\begin{center}\begin{tabular}{|c|c|c|c|c|c|c|c|}\hline
{\sc Super-}&{\sc Represen-}&{\sc Transfor-}&{\sc
Decom-}&\multicolumn{4}{|c|}{\sc Global}\\
{\sc fields}&{\sc tations}&{\sc mations}&{\sc
positions}&\multicolumn{4}{|c|}{\sc Symmetries}\\\cline{5-8}
\cline{5-8}
&{\sc under $G_{\rm LR}$}&{\sc under $G_{\rm LR}$}&{\sc under
$G_{\rm SM}$}& $R$ &{PQ} &{$B$}&{$D$}
\\\hline
\multicolumn{8}{|c|}{\sc Matter Fields}\\\hline
{$l_i$} &{$({\bf 1, 2, 1}, -1)$}&$l_i U_{\rm L}^\tr$&$l_i({\bf 1,2},-1/2)$& $0$ & $-2$ &$0$&$0$\\
{$l^c_i$} & {$({\bf 1, 1, 2}, 1)$}&$U_{\rm R}^\ast l^c_i$& $\nu_i^c({\bf 1, 1}, 0)$&$2$&{$0$}&{$0$}&$0$\\
&&&$e_i^c({\bf 1, 1}, 1)$&&&&\\
{$q_i$} &{$({\bf 3, 2, 1}, 1/3)$}&$q_i U_{\rm L}^\tr\ U_{\rm
C}^\tr$&$q_{i}({\bf 3,2},1/6)$& $1$ & $-1$ &$1/3$&$0$\\
{$q^c_i$} & {$({\bf \bar 3, 1, 2},-1/3)$}&$U_{\rm C}^\ast\ U_{\rm
R}^\ast q^c_i$& $u^c_i ({\bf \bar 3, 1}, -2/3)$&$1$ &{$-1$}&$-1/3$&$0$ \\
&&& $d_i^c ({\bf \bar 3, 1}, 1/3)$&&&&\\\hline
\multicolumn{8}{|c|}{\sc Extra Matter Fields} \\ \hline
$\bar D_{\rm a}$&{$({\bf \bar 3, 1, 1}, 2/3)$}&$U_{\rm C}^\ast
D_{\rm a}$&$\bar D_{\rm a} ({\bf \bar 3, 1},-1/3)$& $2$ & $1$&$0$&$-1$\\
$D_{\rm a}$&{$({\bf 3, 1, 1}, -2/3)$}&$D_{\rm a}U_{\rm
C}^\tr$&$D_{\rm a} ({\bf 3, 1},1/3)$& $2$ & $1$&$0$&$1$\\
$H_{\rm a}$&$({\bf 1, 2, 2}, 0)$ &$U_{\rm L}H_{\rm a} U_{\rm
R}^\tr$&$h_{\rm a}({\bf 1, 2}, 1/2)$& $2$ & $1$ &$0$&$0$\\
&&&$\bar h_{\rm a}({\bf 1, 2}, -1/2)$& &&&
\\ \hline
\multicolumn{8}{|c|}{\sc Higgs Fields}
\\\hline
{$S$} & {$({\bf 1, 1, 1}, 0)$}&$S$&$S~({\bf 1, 1}, 0)$&$4$ &$0$ &$0$&$0$ \\
{$\bar \Phi$}&$({\bf 1, 1, 2}, -1)$&$\bar\Phi U_{\rm R}^\tr$&
$\bar \nu_{\Phi}^c({\bf 1, 1}, 0)$&{$0$}&{$0$}&{$0$}&$0$\\
&&&$\bar e^c_{\Phi}~({\bf 1, 1}, -1)$&&&&\\
{$\Phi$} &{$({\bf 1, 1, 2}, 1)$}&$U_{\rm R}^\ast \Phi $&
$\nu_{\Phi}^c({\bf 1, 1}, 0)$&{$0$}&{$0$} & {$0$}&$0$ \\
&&&$e^c_\Phi~({\bf 1, 1}, 1)$&&&&\\
\hline
{$P$} & {$({\bf 1, 1, 1}, 0)$}&$P$&$P~({\bf 1, 1}, 0)$&$4$ &$0$ &$0$&$0$\\
{$\bar Q$}&$({\bf 1, 1, 1}, 0)$&$\bar Q$&$\bar Q~({\bf 1, 1}, 0)$& {$0$}&{$-2$}&{$0$}&$0$\\
{$Q$} &{$({\bf 1, 1, 1}, 0)$}&$Q$&$Q~({\bf 1, 1}, 0)$& {$0$}&{$2$}
& {$0$}&$0$ \\ \hline
{$h$} & {$({\bf 1, 2, 2}, 0)$}&$U_{\rm L}h U_{\rm R}^\tr$&
$H_u~({\bf 1, 2}, 1/2)$&$2$ &$2$&$0$&$0$\\
& &&$H_d~({\bf 1, 2}, -1/2)$&&&&\\ \hline
\end{tabular}
\end{center}
\caption{\sl The representations, the transformations under
$G_{\rm LR}$, the decompositions under $G_{\rm SM}$ as well as the
extra global charges of the superfields of our model. Here,
$U_{\rm C}\in SU(3)_{\rm C},~U_{\rm L}\in SU(2)_{\rm L},~U_{\rm
R}\in SU(2)_{\rm R}$ and $\tr$ and $\ast$ stand for the transpose
and the complex conjugate of a matrix respectively.}\label{tabLR}
\end{table}
\renewcommand{\arraystretch}{1.0}

In particular, the superpotential, $W$, of our model reads:
\beq\label{Wtot} W= W_{\rm FHI}+W_{\rm PQ}+ \lambda S\bar Q Q\ +\
W_{\rm MSSM}+W_{\rm DW}, \eeq
where $W_{\rm FHI}$ and $W_{\rm PQ}$ are given by \eqs{Whi}{Wpq}
respectively and the anticipated in \Sref{plan} unavoidable
coupling is included. In addition,

$\bullet$ $W_{\rm MSSM}$ is the part of $W$ which contains the
usual terms of the \emph{Minimal SUSY SM} (MSSM), supplemented by
a mass term and Yukawa interactions for right-handed neutrinos,
$\nu^c_i$:
\beq \label{Wmssm} W_{\rm MSSM}=\lambda_\mu {\bar Q^2h^2\over
2\mP}+y_{\nu ij} {\bar \Phi l^c_i \bar \Phi l^c_j\over m_{\rm
P}}+y_{l ij} l_ihl^c_j  + y_{q ij} q_ihq^c_j\>. \eeq
Here, the $i$th generation $SU(2)_{\rm L}$ doublet LH quarks and
leptons are denoted by $q_i$ and $l_i$ respectively, whereas the
$SU(2)_{\rm R}$ doublet antiquarks and antileptons by $q^c_i$ and
$l^c_i$ respectively. The electroweak Higgs are contained in a
$SU(2)_{\rm L}\times SU(2)_{\rm R}$ bidoublet Higgs $h$. The first
term in the \emph{right-hand side} (RHS) of Eq.~(\ref{Wmssm})
generates the $\mu$ term of MSSM via the PQ breaking scale -- see
\Sref{pqpt} --, while the second term generates intermediate scale
masses for $\nu^c_i$  and, thus, seesaw masses \cite{susyhybrid}
for the light neutrinos -- the coupling constant matrix $y_{\nu
ij}$ is considered diagonal.

$\bullet$ $W_{\rm DW}$ is the part of $W$ which gives intermediate
scale masses via $\vev{\bar Q}$ -- see \Sref{pqpt} -- to $\bar
D_{\rm a}-D_{\rm a}$ and $H_{\rm a}$. Namely,
\beq\label{Wdw} W_{\rm DW} =\lambda_{D\rm a}\bar Q\bar D_{\rm a}
D_{\rm a}+\lambda_{H\rm a}\bar Q H_{\rm a}^2, \eeq
where the coupling constant matrices $\lambda_{D\rm a}$ and
$\lambda_{h\rm a}$ are considered diagonal. Although these matter
fields acquire intermediate scale masses after the PQ breaking,
the unification of the MSSM gauge coupling constants is not
disrupted at one loop. In fact, if we estimate the contribution of
$\bar D_{\rm a}, D_{\rm a},$ and $\bar H_{\rm a}$ to the
coefficients $b_1,~b_2,$ and $b_3$, controlling \cite{Jones} the
one loop evolution of the three gauge coupling constants $g_1,
g_2,$ and $g_3$, we find that the quantities $b_2-b_1$ and
$b_3-b_2$ (which are \cite{Jones} crucial for the unification of
$g_1, g_2,$ and $g_3$) remain unaltered.

The K\"ahler potential for our model can include interference
terms of $S$ and $P$ even at the quadratic level, i.e., it has the
form
\bea \nonumber K&=&|S|^2+|P|^2+\ck (S P^*+ S^*P)+
b{|S|^4\over4\mP^2}+c{|P|^4\over4\mP^2}+ d{|S|^2|P|^2\over\mP^2}+\
{ e|S|^2+ f|P|^2\over2\mP^2}\lf S P^*+ S^*P\rg\nonumber\\&+&\
{{\rm g}\over4\mP^2}\left[\lf S P^*\rg^2+ \lf S^*P\rg^2\right]+\
\cdots,\label{minK}\eea
where all the coefficients $\ck, b,c,d,e,f$ and ${\rm g}$ are
taken, for simplicity, real. The ellipsis represents terms
involving the waterfall fields ($\Phi$, $\bar \Phi$, $Q$, and
$\bar Q$) which have negligible impact on our analysis.

\subsection{The Cosmological Scenario}\label{m2}

The F--term SUGRA scalar potential, $V_{\rm SUGRA}$ of our model
can be found by applying the well-known formula -- see e.g.
Ref.~\cite{hybrid}:
\begin{equation}
V_{\rm SUGRA}=e^{K/\mP^2}\left(F_{i^*}^*\lf K_{,ji^*}\rg^{-1}
F_j-3\frac{\vert W\vert^2}{\mP^2}\right)~~\mbox{with}~~F_i=W_{,i}
+K_{,i}{W\over\mP^2}\cdot \label{sugra}
\end{equation}
Here,  a subscript $,i~[,i^*]$ denotes derivation with respect to
the complex scalar field $i~[i\,^{*}]$. Taking the limit
$\mP\to\infty$, we can obtain the SUSY limit of $V_{\rm SUGRA}$,
$V_{\rm F}$, which turns out to be
\bea \label{VF} V_{\rm F} &=&
{1\over(1-\ck^2)}\left(\left|\kappa\left(\bar\Phi\Phi-M^2\right)+\lambda
\bar Q Q\right|^2+\ \kappa_a^2\left|\bar Q Q-M^2_a\right|^2\right)
+\ \kappa^2|S|^2\left(|\bar\Phi|^2+ |\Phi|^2\right)\nonumber\\&+&
\left|\lambda  S Q+\kappa_a P Q+ \lambda_{D\rm a}\bar D_{\rm a}
D_{\rm a}+\lambda_{H\rm a}H_{\rm a}^2 \right|^2 +\ |\bar
Q|^2\left(\left|\lambda  S +\kappa_a P \right|^2+ \lambda_{D\rm
a}^2\left(|\bar D_{\rm a}|^2+ |D_{\rm a}|^2\right)+ \lambda_{H\rm
a}^2|H_{\rm a}|^2\right) \nonumber\\ &-&
{\ck\over(1-\ck^2)}\Big[{\kappa_a}\lf \bar Q^*
Q^*-M^2_a\rg\big[\kappa\lf\bar\Phi\Phi-M^2\right) +\ \lambda \bar
Q Q\big]+ \mbox{c.c.}\Big],\eea
where the complex scalar components of the superfields are denoted
by the same symbol. From the potential in Eq.~(\ref{VF}) and
taking into account that $M\gg f_a$, we find that the SUSY vacuum
lies at the directions -- cf. \eqs{fhivev}{PQvev}:
\beqs\bea && \vev{S}\simeq0,~\vev{P}\simeq0,~\vev{\bar
D_a}=\vev{D_a}=\vev{H_a}=\vev{\seHb}=\vev{\seH}=0,\label{vevs0}\\
&&\vev{\snHb}=\vev{\snH}=M~~\mbox{and}~~\vev{\phi_Q}={f_a}\>,\label{vevs}\eea\eeqs
where we have introduced the  canonically normalized scalar field
$\phi_Q=2Q=2\bar Q$ -- cf. \Eref{PQvev}. As a consequence, $W$
leads to a spontaneous breaking of $G_{\rm LR}$ and ${U}(1)_{\rm
PQ}$. In addition, $W$ gives rise to a stage of FHI and a PQPT,
since $V_{\rm F}$ possesses \emph{two} D-- and F--flat directions
for
%
\beqs \bea \label{flat} &&~\snHb=\snH=\seHb=\seH=\bar Q= Q=\bar
D_a= D_a=H_a=0 \\~~\mbox{and}~~&&~S=\seHb=\seH=\bar Q=
Q=0~~\mbox{and}~~\snHb=\snH=M, \label{flat1}\eea \eeqs
with a constant potential energy density respectively
\beq \label{V0} {\ftn\sf
(a)}~~\Vhio\simeq{\kappa^2M^4}/{(1-\ck^2)}~~~~\mbox{and}~~~~
{\ftn\sf (b)}~~\Vpqo=\kappa_a^2f_a^4/16.\eeq
By constructing the scalar spectrum along the direction of
\Eref{flat} -- see \Tref{tab} --, we can deduce that it can be
used as inflationary path since it corresponds to a classically
flat valley of minima for
\beq\label{stab}{\sf
(a)}~~|S|>{M\over\sqrt{1-\ck^2}}~~~\mbox{and}~~~{\sf
(b)}~~|\sigma_a|>
{\sqrt{\kappa(\lambda-\ck\ka)\over1-\ck^2}}M,~~\mbox{where}~~
\sigma_a=\lambda S+\kappa_aP.\eeq
Since $\Vpqo\ll\Vhio$, $\Vpqo$ can dominate over radiation after
the end of FHI leading to a PQPT. This cosmological scenario can
be attained if \sEref{stab}{a} is violated before \sEref{stab}{b},
since, in this case, we obtain $\snHb=\snH=M~\mbox{and}~\bar Q=
Q=0$ and not $\snHb=\snH=0~\mbox{and}~\bar Q= Q=f_a/2$.

\begin{table}[t]
\renewcommand{\arraystretch}{1.2}
\bec\begin{tabular}{|c||c|c||c|c|}\hline
{\sc Super-} &{\sc Scalars} & {\sc Mass}&{\sc Fermions} & {\sc Mass}\\
{\sc fields} & {\sc (12 real)}&{\sc Squared}&{\sc (6 Weyl)}&{\sc
Squared}\\\hline
$\bar\Phi$, $\Phi$ & $\Re\ \&\ \Im ~[\snHb\pm\snH]$ &
$\kappa^2\left(|S|^2\pm \frac{M^2}{1-\cks^2}\right)$ &
$\psi_{\nu\pm}={\psi_{\bar\nu}\pm
\psi_{\nu}\over\sqrt{2}}$& $\kappa^2|S|^2$ \\
&$\Re\ \&\ \Im~[\seHb\pm\seH]$&&${\psi_{\bar e}\pm\psi_{e}\over\sqrt{2}}$&\\
$\bar Q$, $Q$ & $\Re\ \&\ \Im~[\bar Q\pm Q]$ & $|\sigma_a|^2
\pm\frac{\kappa(\lambda-\cks\ka)M^2}{1-\cks^2}$&
$\psi_{Q\pm}={\psi_{\bar Q}\pm\psi_{Q}\over\sqrt{2}}$ &
$|\sigma_a|^2$\\\hline
\end{tabular}\eec
\caption{\sl The SUSY-breaking mass spectrum along the
inflationary trajectory of Eq.~(2.8{\ftn\sf a}). Here, $\psi_{x}$
with $x=\bar \nu, \nu, \bar e, e, \bar Q$ and $Q$ denote the
chiral fermions associated with the superfields $x=\snHb, \snH,
\seHb, \seH, \bar Q$ and $Q$ respectively.} \label{tab}
\end{table}

\subsection{Evading the Domain-Wall Problem}\label{walls}

Soft SUSY breaking and instanton effects explicitly break
${U(1)_R}\times {U(1)_{\rm PQ}}$ to a discrete subgroup, which can
be found, for every $n$, by solving the system of equations:
\beq \left.\bem
e^{ir R(W)}=1\hfill \cr
e^{ir\sum_i R(i)+p\sum_i PQ(i)}=1\hfill
\cr\eem\right\}\Rightarrow\left\{\bem
4r =0~\lf\mbox{\ftn\sf mod}~2\pi\rg\hfill \cr
-12r +2(n-6)p=0~\lf\mbox{\ftn\sf mod}~2\pi\rg\hfill \cr\eem\right.
~~\mbox{where}~~\left\{\bem
e^{irR}\in U(1)_{R}\hfill \cr
e^{ipPQ}\in U(1)_{\rm PQ},\hfill \cr\eem\right. \eeq
with $r$ [$p$] being a $U(1)_{R}$ [$U(1)_{\rm PQ}$] rotation and
the sum over $i$ is applied over all ${SU(3)_{\rm C}}$ ${\bf 3}$
and ${\bf\bar 3}$ of the model. We conclude that the unbroken
subgroup is $\mathbb{Z}_4\times \mathbb{Z}_{2(n-6)}$. It is then
important to ensure that this subgroup is not spontaneously broken
by $\vev{Q}$ and $\vev{\bar Q}$, i.e., the equations
\beq e^{2i p_s}\vev{Q}=\vev{Q}~~\mbox{and}~~e^{-2ip_s}\vev{\bar
Q}= \vev{\bar Q}~~\Rightarrow~~ 2p_s=0~\lf\mbox{\ftn\sf
mod}~2\pi\rg.\eeq
are satisfied identically -- otherwise, cosmologically disastrous
domain walls are produced \cite{sikivie} at PQPT. This goal can be
accomplished by choosing $n=5$ or $n=7$. Therefore, for these
$n$'s, the domain-wall production during PQPT can be eluded.


\section{The Inflationary Era}\label{fhi}

Below, we describe the salient features of the inflationary
potential in \Sref{fhi1} and  we analyze the inflationary dynamics
in \Sref{fhi2}.

\subsection{The Inflationary Potential} \label{fhi1}
The inflationary potential along the trajectory of \Eref{flat} can
be written as \beq \Vhi=V_{\rm HI0}+V_{\rm HIs}+V_{\rm
HIc},~~\mbox{where}\label{Vhi}\eeq

$\bullet$ $V_{\rm HI0}$ is the dominant contribution to $V_{\rm
HI}$ along the F-flat direction, given in \sEref{V0}{a}.

$\bullet$ $V_{\rm HIs}$ is the SUGRA corrections to $V_{\rm HI}$
which can be found by expanding $V_{\rm SUGRA}$ in
Eq.~(\ref{sugra}) along the trajectory of \Eref{flat}. Namely,
\beqs\bea V_{{\rm HIs}}&\simeq& {\Vhio\over (1-\ck^2)m^2_{\rm
P}}\Big[ A_1|S|^2 + A_{12}\lf S^*P+PS^*\rg +
A_2|P|^2\Big]+{\Vhio\over 4(1-\ck^2)^2m^4_{\rm P}}\Big[ B_1|S|^4 +
B_2|P|^4 \nonumber\\&+&\ B_{3} |S|^2|P|^2+\lf
B_{4}|S|^2+B_{5}|P|^2\rg \lf S^*P+PS^*\rg+B_6\lf \lf S^*P\rg^2+\lf
P^*S\rg^2\rg \Big] \nonumber\\&=& {\Vhio\over2\mP^2}\lf\msp\lf
s^2+q^2\rg+\msm \sg^2\rg+\cdots,\>\>\label{Vhis}\eea
where the coefficients $A_1-A_3$ and $B_1-B_6$, given in
\cref{pqhi}, are functions of the coefficients $\ck,...,{\rm g}$
in \Eref{minK}; the real fields $\sg, s$ and $q$ are the
eigenvectors (corresponding to the eigenvalues $\mspm$) of the
matrix involved in the quadratic part of $V_{\rm HIs}$. This can
be worked out \cite{pqhi} after the quadratic part, $K_{SP}$, of
$K$ in \Eref{minK} has been brought into a canonical form, i.e.,
we obtain also
$$K_{SP}=|S|^2+|P|^2+\ck (S P^*+ S^*P)= \lf\sg^2+s^2+q^2\rg/2.$$
Since $\msm\simeq0$, its corresponding eigenvector, $\sg$, can be
qualified as the inflaton. Note that we need the higher order
terms of $K$ in \Eref{minK} so that we obtain $\msm\leq0$ and
therefore, observationally acceptable $n_{\rm s}$'s -- see
\Sref{cont}. Indeed, for $\ck\neq0$ and $b=c=d=e=f={\rm g}=0$ we
get $\msm=0$.

$\bullet$ $V_{\rm HIc}$ represents the contribution to $V_{\rm
HI}$ from one-loop radiative corrections, due to SUSY-breaking
mass spectrum presented in \Tref{tab}, which can be calculated
\cite{cw} to be
\beq V_{\rm HIc}\simeq{\kappa^2 \Vhio\over 8\pi^2(1-\ck^2)}\left(
\ln {\kappa^2x M^2
\over(1-\ck^2)\Lambda^2}+{3\over2}\right)+{(\lambda-\ck\ka) ^2
\Vhio \over 16\pi^2(1-\ck^2)}\left(\ln {\kappa (\lambda-\ck\ka)
x_a M^2
\over(1-\ck^2)\Lambda^2}+{3\over2}\right)~,\label{Vhic}\eeq\eeqs
with $x=|S|^2(1-\ck^2)/M^2$ and
$x_a=|\sigma_a|^2(1-\ck^2)/\kappa(\lambda-\ck\ka) M^2$. Here, we
take into account that the dimensionality of the representations
to which $\bar{\Phi}$ and $\Phi$ [$\bar Q$ and $Q$] belong is
2~[1] -- see Table~1.

\begin{figure}[t]\vspace*{-0.45cm}
\begin{minipage}{75mm}
\includegraphics[height=3.5in,angle=-90]{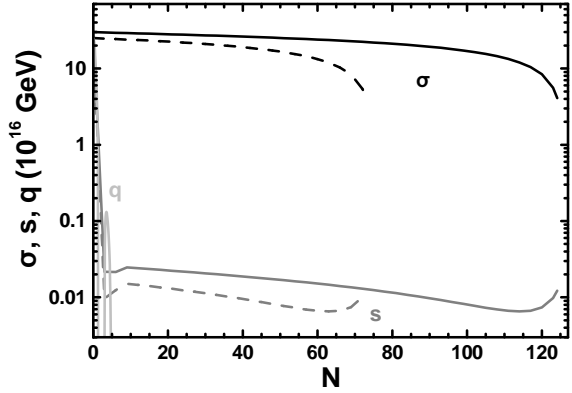}
\end{minipage}
\hfil \hspace*{1.5cm}\begin{minipage}{60mm}
\vspace*{-1.5cm}\caption{\sl The evolution of $\sg$ (black lines),
$s$ (gray lines), and $q$ (light gray lines) as functions of $N$
for the values of the parameters shown in Eq.~(5.4),
$\kp=0.0023,~\ck=-0.011,~b=-0.01$ and $f_{\rm
HIi}=3\cdot10^{17}~\GeV$ (solid lines) or $f_{\rm
HIi}=2.5\cdot10^{17}~\GeV$ (dashed lines). For the employed values
of parameters, the requirements of Sec.~5.1 are
fulfilled.}\label{fig1}
\end{minipage}

\end{figure}


\subsection{The Inflationary Dynamics} \label{fhi2}

The \emph{equations of motion} (e.o.m) of the various fields are
($~~\dot{}=d/dt$ with $t$ the cosmic time):
\beq \ddot f+3H\dot f+V_{{\rm
HI},f}=0~\Rightarrow~H^2f''+3H^2f'+V_{{\rm
HI},f}=0~~\mbox{with}~~f=\sigma,~s,~\mbox{and}~q~\eeq
and $' = d/ dN$ where $N=\ln \left(R/R_{\rm HIi}\right)$. Here
$R(t)$ is the scale factor of the universe and the subscript
``HIi'' denotes values at the onset of FHI. We impose the
following initial conditions (at $N=0$):
\beq f_{\rm HIi}=f(0)=(1.5-4.5)\cdot10^{17}~\GeV~~\mbox{and}~~
f'(0)=0~~\mbox{with}~~f=\sigma,~s,~\mbox{or}~q.\eeq
When $f_{\rm HIi}$ is large enough, $s$ reaches an attractor and
our results are independent of the precise value of $f_{\rm HIi}$,
as can be clearly deduced from \Fref{fig1}, where we plot $\sg$
(black lines), $s$ (gray lines), and $q$ (light gray lines) as
functions of $N$ for $f_{\rm HIi}=3\cdot10^{17}~\GeV$ (solid
lines) or $f_{\rm HIi}=2.5\cdot10^{17}~\GeV$ (dashed lines). In
both cases, we adopt the values of the parameters shown in
\Eref{para2}, $\kp=0.0045,~b=-0.01$ and $\ck=-0.011$ which fulfill
the requirements of \Sref{cont1}. For both choices of $f_{\rm
HIi}$'s, we obtain $m_-^2=-0.0126$, $m_+^2=1.83$, $\ns=0.968$,
$N_{\rm HI*}=52$, $\sg_{\rm HIf}=4.08\cdot10^{16}~\GeV$ and
$s_{\rm HIf}=1.3\cdot10^{14}~\GeV$ although in the first [second]
case we obtain $N_{\rm HI}=87.1$ [$N_{\rm HI}=175.5$] -- $N_{\rm
HI}$ and $N_{\rm HI*}$ are defined below \Eref{Nhi} in
\Sref{cont1}. We observe that immediately after the onset of FHI,
$q$ decreases sharply, whereas the value of $s$ at the end of FHI,
$s_{\rm HIf}$, turns out to be just mildly, and not drastically
reduced compared to $\sigma$ -- in sharp contrast to the situation
of Ref.~\cite{kawasaki}. This is due to the participation of $s$
in both \eqs{Vhis}{Vhic}.

\section{The Post-Inflationary Era} \label{pfhi}

We below describe the post-inflationary evolution of our model,
presenting the dynamics of the two fields, $\sg$ and $s$, in
\Sref{pfhi1} and this of the two reheating processes in
\Sref{pfhi2}. For later convenience, we arrange in \Tref{tab3} the
mass spectrum of our model at the SUSY vacuum of
\eqs{vevs0}{vevs}.

\renewcommand{\arraystretch}{1.4}
\newcommand{\ldu}{\ensuremath{\uplambda}}

\begin{table}[!t]
\bec\begin{tabular}{|c|c|c||c|c|c|}\hline
\multicolumn{2}{|c|}{\sc Eigenstates}&\multicolumn{1}{c||}{\sc
Eigenvalues}& \multicolumn{2}{c|}{\sc
Eigenstates}&\multicolumn{1}{c|}{\sc
Eigenvalues}\\\cline{1-2}\cline{4-5}
{\sc Bosons} &{\sc Fermions} & (\sc Masses)&{\sc Bosons} &{\sc Fermions} & {\sc (Masses)}\\
\hline
$S,{\delta\snHb+\delta
\snH\over\sqrt{2}}$&${\psi_{S}\pm\psi_{\nu+}\over\sqrt{2}}$ &
$m_{\rm I}=\sqrt{2}\kp M$& $P,{\delta\bar Q+\delta Q\over
\sqrt{2}}$&${\psi_{P}\pm\psi_{Q+}\over\sqrt{2}}$
& $m_{\rm PQ}=\ld_a {f_a\over\sqrt{2}}$\\
$\Re[\snHb-\snH]$&$ {\psi_{\bar\nu}-\psi_{\nu}\over\sqrt{2}}$ &
$\sqrt{5/2}gM$&${\delta\bar Q-\delta Q\over\sqrt{2}}$&${\psi_{\bar Q}-\psi_{Q}\over\sqrt{2}}$ & $0$\\
$\bem  \Re[\seHb-\seH] \cr {\sf\ftn Im} {\ftn [\bar e^c_{\Phi}+e^c_{\Phi}]}\cr\eem $
&$\psi_{e},\psi_{\bar e}$ & $gM$&$\tilde\nu^c_i$&$\nu^c_i$ & $2y_{\nu i}M^2/\mP$\\
${A^1_{\rm R}\pm iA^2_{\rm R}\over \sqrt{2}}$&${\ldu^1_{\rm R}\pm
i\ldu^2_{\rm R}\over \sqrt{2}}$& $gM$&$D_{{\rm a}},~\bar D_{{\rm
a}}$&$\psi_{D_{{\rm a}}},~\psi_{\bar
D_{{\rm a}}}$& $\ld_{D{\rm a}}f_a/2$\\
${A^3_{\rm R}\pm A_{B-L}\over \sqrt{2}}$&${\ldu^3_{\rm R}\pm
\ldu_{B-L}\over \sqrt{2}}$ & $0,\sqrt{5/2}gM$&
$H_{{\rm a}}$&$\psi_{H_{{\rm a}}}$ & $\ld_{H{\rm a}}f_a/2$\\
\hline
\end{tabular}\eec
\caption{\sl The mass spectrum of the model at the SUSY vacuum of
Eq.~(2.7{\ftn\sf a}) and (2.7{\ftn\sf b}). Here, $A_{\rm R}^m$
$[\ldu_{\rm R}^m]$ with $m=1,2,3$ are gauge bosons [gauginos]
associated with the $SU(2)_{\rm R}$, while these corresponding to
$U(1)_{B-L}$ are denoted by $A_{B-L}$ $[\ldu_{B-L}]$. Also,
$\psi_{x}$ with $x=S,\bar \nu, \nu, \bar e, e, P, \bar Q, Q, \bar
D_{\rm a}, D_{\rm a}$ and $H_{\rm a}$ denote the Weyl spinors
associated with the superfields $x=S, \snHb, \snH, \seHb,\seH, P,
\bar Q, Q, \bar D_{\rm a}, D_{\rm a}$ and $H_{\rm a}$
respectively.} \label{tab3}
\end{table}

\subsection{The Dynamics of Scalars} \label{pfhi1}

When FHI is over, the inflaton system with mass $m_{\rm I}$ -- see
\Tref{tab3} -- consisting of the two complex scalar fields $S$ and
$(\delta\snH-\delta\snHb)/\sqrt{2}$ -- where $\delta\snH= \snH -M$
and $\delta\snHb= \snHb -M$ -- settles into a phase of damped
oscillations and decays reheating the universe to a temperature
\beq T_{\rm 1rh}=\left(72\over5\pi^2g_{{\rm 1rh}*}\right)^{1/4}
\sqrt{\Gamma_1 m_{\rm
P}},~~~~\mbox{where}~~~~\Gamma_1={1\over16\pi}\lambda ^2\,m_{\rm
I}\label{T1rh} \eeq is the decay width emerging from the third
term in the RHS of \Eref{Wtot}. Here, $g_{\rm
1rh*}\simeq438.75$~[$g_{\rm 1rh*}\simeq513.75$] for $n=5$ [$n=7$]
counts the relativistic degrees of freedom of the model.

For $\ld\simeq(0.05-0.1)$, we get $T_{\rm 1rh}>V_{\rm PQ0}^{1/4}$.
Therefore, we obtain \emph{matter domination} (MD) for $T\geq
T_{\rm 1rh}$ and \emph{radiation domination} (RD) for
$\Vpqo^{1/4}\lesssim T\lesssim T_{\rm 1rh}$. During MD, $s$
\cite{kawasaki, moroiHm} acquires an effective mass equal to
$\sqrt{3/2}H$. Solving its e.o.m for $N>N_{\rm HI}$, we can
extract its value, $s_{\rm PQi}$, -- and the corresponding value
of $P$, $P_{\rm PQi}$ -- at $T=T_{\rm 1rh}$ which coincides with
its value at the onset of PQPT since, during the subsequent RD
era, $s$ remains \cite{kawasaki, moroiHm} frozen. Namely we find
\beq P_{\rm PQi}=A_P s_{\rm PQi}\>\>\>\>\mbox{with}\>\>\>\>s_{\rm
PQi}\simeq\left({\rho_{\rm 1rh}\over\Vhio}\right)^{1/4}s_{\rm
HIf}\>\>\>\>\mbox{and}\>\>\>\>\rho_{\rm 1rh}={\pi^2\over30}g_{\rm
1rh*}T_{\rm 1rh}^4,\label{spqi}\eeq
where $A_P$ is a function \cite{pqhi} of the coefficients of $K$
in \Eref{minK}.

For $T\lesssim\Vpqo^{1/4}$, $W$ in \Eref{Wtot} is dominated by
$W_{\rm PQ}$ in \Eref{Wpq} and the relevant F-term scalar
potential is given in \Eref{Vpqf} which along the flat direction
of \Eref{flat1} gives rise to the constant potential energy
density of \sEref{V0}{b}. Assuming gravity mediated soft SUSY
breaking, the potential along the direction of \Eref{flat1} for
$|P|\geq f_a/2$ has the form:
\beq \label{Vpq} \Vpq\simeq\Vpqo+m^2_P\,|P|^2-\sqrt{2\Vpqo}\;|{\rm
a}_P| |P|+{\kappa_a^2 \Vpqo\over 16\pi^2}\left(\ln
{\kappa_a^2|P|^2\over\Lambda^2}+{3\over2}\right),\eeq
where the 2nd and 3rd contributions arise from soft SUSY breaking
effects and the forth contribution represents the 1-loop
corrections \cite{cw} due to the SUSY breaking \cite{pqhi}. Mainly
due to this last contribution, $V_{\rm PQ}$ does not give rise to
another FHI, since the $\eta$-criterion is spoiled. Nonetheless,
when $|P| < f_a/2$, an instability occurs along the $|P|$-axis
triggering thereby a PQPT. If, in addition, $|P_{\rm PQi}| >
f_a/2$ we obtain an out-of-equilibrium decay of the PQ system,
i.e., a secondary reheating.

During this latter phase, the PQ system with mass $m_{\rm PQ}$ --
see \Tref{tab3} -- comprised of the complex fields $P$ and
$(\delta\bar Q+\delta Q)/\sqrt{2}$ -- where $\delta Q= Q -f_a/2$
and $\delta\bar Q= \bar Q -f_a/2$ -- enters a phase of
oscillations reheating the universe to the temperature
\beq T_{\rm 2rh}=\left(72\over5\pi^2g_{\rm 2rh*}\right)^{1/4}
\sqrt{\Gamma_{2} m_{\rm
P}},~~~~\mbox{where}~~~~\Gamma_{2}={1\over2\pi}\lambda_\mu^2\left({f_a\over
2m_{\rm P}}\right)^2m_{\rm PQ}\label{T2rh}\eeq is the decay width
emerging from the first term in the RHS of \Eref{Wmssm}. Also,
$g_{\rm 2rh*}=232.5$ counts the relativistic degrees of freedom of
MSSM plus the content of the axion supermultiplet.

\subsection{The Dynamics of Reheating Processes}\label{pfhi2}

A more accurate description of the reheating dynamics can be
obtained by solving the relevant Boltzmann equations. In
particular, the energy density, $\rho_1$ [$\rho_2$], of the
oscillatory system which reheats the universe at the temperature
$T_{1\rm rh}$ [$T_{2\rm rh}$], the energy density of produced
radiation, $\rho_{\rm R}$, and the number density of $\Gr$,
$n_{\Gr}$, satisfy the equations \cite{pqhi}:
\beq \left.\bem
\dot \rho_1+3H\rho_1+\Gamma_1 \rho_1=0,\hfill \cr
\dot\rho_2+3H\rho_2+\Gamma_2\rho_2=0,\hfill \cr
\dot\rho_{\rm R}+4H\rho_{\rm R}-
\Gamma_1\rho_1-\Gamma_2\rho_2=0,\hfill \cr
\dot n_{\Gr}+3Hn_{\Gr}-C_{\Gr} \lf n^{\rm eq}\rg^2=0,\hfill
\cr\eem\right\}~~~~\mbox{with}~~~~\left\{\bem H=\left(\rho_1
+\rho_2+\rho_{\rm R} \right)^{1/2}/\sqrt{3}\mP,\hfill \cr C_{\Gr}
= 3\pi\sum_{i=1}^{3} c_i g_i^2 \ln\left({k_{i}/
g_i}\right)/16\zeta(3)\mP^2,\hfill \cr n^{\rm
eq}={\zeta(3)T^3/\pi^2},~T=30\rho_{\rm R}/g_*\pi^2. \hfill
\cr\eem\right.\label{Boltz}\eeq
Here, $(k_i)=(1.634,1.312,1.271)$, $(c_i)=(33/5,27,72)$ and  $g_*
(T)=g_{1\rm rh*}$ [$g_* (T)=g_{2\rm rh*}$] for $T\geq T_{\rm PQ}$
[$T<T_{\rm PQ}$] where $T_{\rm PQ}$ is defined as the solution of
the equation $\rho_{\rm R}\lf T_{\rm PQ}\rg=\Vpqo$. We use the
following initial conditions -- the quantities below are
considered as functions of the independent variable $\bar
N=\ln\left(R/R_{\rm HIf}\right)$ with $R_{\rm HIf}$ being the
value of the scale factor at the end of FHI:
\beq \rho_1(0)=\Vhio,~\rho_{\rm R}(0)=n_{\Gr}(0)=0,~\mbox{and}~
\rho_2(\bar N_{\rm PQ})=\Vpqo,\label{inic} \eeq
where $\bar N_{\rm PQ}$ is the value of $\bar N$ corresponding to
the temperature $T_{\rm PQ}$.

\begin{figure}[t]\vspace*{-0.45cm}
\begin{minipage}{75mm}
\includegraphics[height=3.5in,angle=-90]{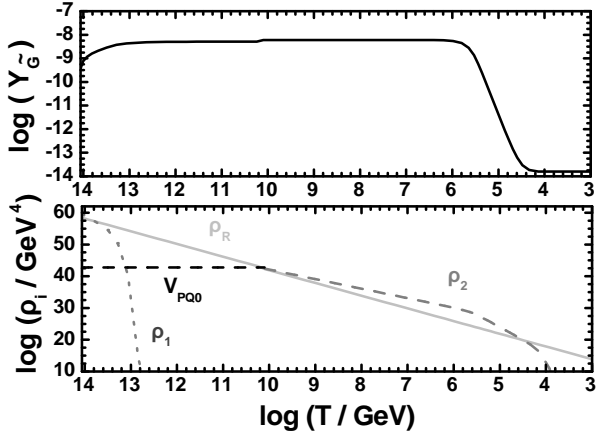}
\end{minipage}
\hfil \hspace*{1.5cm}\begin{minipage}{60mm}
\vspace*{-1.1cm}\caption{\sl The evolution of the quantities
$\log\rho_i$ with $i=1$ (gray dotted line), $i=2$ (gray dashed
line), $i={\rm R}$ (gray line), $\log\Vpqo$ (black dashed line),
and $\log Y_{\Gr}$ (black solid line) as functions of $\log T$ for
$\kp=0.0023, \ck=-0.011,~b=-0.01$ and the values of the remaining
parameters shown in Eq.~(5.4). For the employed vales of
parameters, the requirements of Sec.~5.1 are
fulfilled.}\label{fig2}
\end{minipage}

\end{figure}

In Fig.~\ref{fig2}, we illustrate the cosmological evolution of
the quantities $\log\rho_i$ with $i=1$ (dotted gray line), $i=2$
(dashed gray line), and $i={\rm R}$ (gray line), $\log\Vpqo$
(black dashed line), and $\log Y_{\Gr}$ (black solid line) as
functions of $\log T$ for the values of the parameters adopted in
\Fref{fig1}. We observe that FHI is followed successively by a MD
era, which lasts until $T=6\cdot10^{13}~\GeV\simeq T_{1\rm rh}$
(where $\rho_1=\rho_{\rm R}$), a RD epoch, terminated at $T_{\rm
PQ}=1.4\cdot10^{10}~\GeV$, a MD era, completed at
$T=3.1\cdot10^4~\GeV\simeq T_{2\rm rh}$ (where $\rho_2=\rho_{\rm
R}$) and followed by the conventional RD epoch. We also see that
the $\Gr$ abundance immediately after FHI is
$Y_{1\Gr}=5.5\cdot10^{-9}$ which can be estimated by \cite{brand,
kohri}
\beq Y_{1\Gr}={n_{\Gr}\over{\sf
s}}(\Trha)\simeq1.9\cdot10^{-12}\left({\Trha\over10^{10}~{\rm
GeV}}\right)~~\mbox{with}~~{\sf
s}={2\pi^2\over45}g_*T^3.\label{Y1gr} \eeq However, the $\Gr$
abundance decreases sharply to $Y_{2\Gr}=1.6\cdot10^{-14}$ which
can be approximated by \beq \label{Y2gr}
Y_{2\Gr}={n_{\Gr}\over{\sf s}}(\Trhb)\simeq\lf{\pi^2\over
30}g_{\rm 1rh*}\rg^{1/4}{\Trhb\over\Vpqo^{1/4}}Y_{1\Gr}.\eeq We
observe that $Y_{2\Gr}$ is suppressed relative to $Y_{1\Gr}$ by
the ratio $\Trhb/\Vpqo^{1/4}\ll1$ due to the entropy released
during the out-of-equilibrium decay of the PQ system.
Interestingly enough, the dilution of $Y_{1\Gr}$ is independent of
$\ka$ -- see Eqs.~(\ref{V0}{b}) and (\ref{T2rh}).

\section{Testing Against Observations}\label{cont}

We below exhibit the constraints that we impose on our
cosmological set-up in \Sref{cont1} and delineate the allowed
parameter space of our model in Sec.~\ref{num}.

\subsection{Observational Constraints}\label{cont1}

The parameters of our model can be restricted imposing the
following requirements -- note that in the point (v) below we
adopt an updated, compared to our analysis in \cref{pqhi}, version
of the relevant constraint :

\begin{itemize}

\item[{\bf (i)}] The violation of the instability conditions in
\Eref{stab} occurs according to the desired order.

\item[{\bf (ii)}] The number of $e$-foldings $N_{\rm HI*}$ that
the scale $k_*=0.002/{\rm Mpc}$ suffered during FHI has to be
sufficient to resolve the horizon and flatness problems of
Standard Big Bang cosmology:
\beq N_{\rm HI*}=N_{\rm HI}-N_*\simeq23+{2\over 3}\ln{V^{1/4}_{\rm
HI0}\over{1~{\rm GeV}}}-{1\over 3}\ln{V^{1/4}_{\rm
PQ0}\over{1~{\rm GeV}}}+ {1\over3}\ln {T_{\rm 1rh}T_{\rm
2rh}\over{1~{\rm GeV}^2}}\>,\label{Nhi}\eeq
where $N_*$ and $N_{\rm HI}$ are the values of $N$ from the onset
of FHI until $k_*$ crossed outside the horizon of FHI and the end
of FHI, respectively. $N_{\rm HI}$ is the largest $N$ at which we
obtain violation of \sEref{stab}{a} or of the condition: \beq{\sf
max}\{\epsilon(\sigma(N)),|\eta(\sigma(N))|\}\leq1,~~\mbox{with}~~\epsilon\simeq{m^2_{\rm
P}\over2}\left(\frac{V_{\rm HI,\sg}}{V_{\rm
HI}}\right)^2~~\mbox{and}~~\eta\simeq m^2_{\rm P}~\frac{V_{\rm
HI,\sg\sg}}{V_{\rm HI}}\>\cdot\label{sr}\eeq

\item[{\bf (iii)}] The power spectrum of the curvature
perturbation at $k=k_{*}$ is to be confronted with the WMAP7 data:
\beq \Delta_{\cal R*}=\left.V_{\rm HI}^{3/2}\over{2\sqrt{3}\, \pi
m^3_{\rm P}}|V_{{\rm HI},\sigma}
|\right\vert_{N=N_*}\simeq4.93\cdot 10^{-5}.\label{Pr}\eeq

\item[{\bf (iv)}] The mass, $gM$, of the lightest gauge boson at
the SUSY vacuum -- see Table~3 -- is to take the value dictated by
the unification of the gauge coupling constants within MSSM, i.e.,
\beq \label{Mgut} {g M}\simeq2 \cdot
10^{16}~\GeV\>\Rightarrow\>M\simeq2.86\cdot
10^{16}~\GeV~~\mbox{with}~~g\simeq0.7,\eeq
being the value of the unified gauge coupling constant - not to be
confused with the coefficient ${\rm g}$ appearing in \Eref{minK}.
Note that $G_{\rm LR}$ is considered embedded in the $SO(10)$.

\item[{\bf (v)}] The spectral index, $n_{\rm s}$, is to be
consistent with the fitting of the WMAP7 results by the
$\Lambda$CDM model (with negligible running $\alpha_{\rm
s}\simeq0$), i.e.,
\beq n_{\rm s}=1-6\epsilon(N_*)\ +\
2\eta(N_*)=0.968\pm0.024~\Rightarrow~0.944\lesssim n_{\rm s}
\lesssim 0.992~~\mbox{at 95\% c.l.}\label{nsw}\eeq

\item[{\bf (vi)}] In order for the PQPT to take place after a
short temporary domination of $\Vpqo$, we require:
\beq \left|P_{\rm PQi}\right|>f_a/2 ~\Rightarrow~s_{\rm
PQi}>f_a/A_p. \label{Ppqi}\eeq

\item[{\bf (vii)}] Assuming unstable $\Gr$, we impose an upper
bound on $Y_{2\Gr}$ in order to avoid problems with the standard
Big Bang nucleosynthesis \cite{kohri}:
\beq Y_{2\Gr}\lesssim\left\{\bem
10^{-14}\hfill \cr
10^{-13}\hfill \cr\eem
\right.~~~~\mbox{for \Gr\ mass}~~~~m_{\Gr}\simeq\left\{\bem
0.69~{\rm TeV}\hfill \cr
10.6~{\rm TeV.}\hfill \cr\eem
\right.\label{Ygr}\eeq\end{itemize}

\subsection{Numerical Results}\label{num}

As can be seen from the analysis above, our cosmological set-up
depends on the following parameters: $
\kp,~\ka,~\ld,~f_a,~\lambda_\mu,~n,~\ck,~b,~c,~d,~e,~f,~\mbox{and
g}. $ We fix throughout our computation: \beq
\ld=0.1,~f_a=10^{12}~\GeV,~\ka=\lambda_\mu=0.01,~n=5~~\mbox{and}~~c=d=e=f={\rm
g}=0.1.\label{para2}\eeq
The chosen $f_a$ and $\lambda_\mu$ result to $\mu\simeq1~{\rm
TeV}$ via the first term of the RHS of \Eref{Wmssm}. Also, the
selected $\ka$ and $\ld$ play a crucial role in the determination
of $T_{\rm 1rh}$ and $T_{\rm 2rh}$ -- via Eq.~(\ref{T1rh}) and
(\ref{T2rh}) and facilitate the violation of the conditions in
\Eref{stab} in the desired order. Their variation, thought, does
not cause drastic changes in the inflationary predictions. The
same is also valid for the fixed in \Eref{para2} parameters of
$K$, in \Eref{minK} which -- contrary to $\ck$ and $b$ -- do not
influence the computation of $\msp$ and $\msm$. As we show below,
the selected values above give us a wide and natural allowed
region of the remaining fundamental inflationary parameters
($\kp,\ck,$ and $b$).

Besides the parameters above, in our computation, we use as input
parameters the quantities $N_*$ and $f_{\rm HIi}$ with $f=\sg,~s,$
and $q$. We set $f_{\rm HIi}\simeq(1.5-3.5)\cdot10^{17}~\GeV$ so
as to obtain $N_{\rm HI}\simeq70-140$. We then restrict $M$ and
$N_*$ so that Eqs.~(\ref{Nhi}) and (\ref{Pr}) are fulfilled. It is
gratifying that our model supports solutions which simultaneously
fulfill \eqs{Pr}{Mgut} contrary to most realizations of FHI -- cf.
\cref{gpp} -- which requires, via \Eref{Pr}, $M$'s lower than
those indicated in \Eref{Mgut}. We finally check if the preferred
hierarchy in the violation of Eqs.~(\ref{stab}{a}) and
(\ref{stab}{b}) is achieved and proceed imposing the requirements
(v) - (vii) of \Sref{cont1}.

\begin{figure}[t]\vspace*{-0.45cm}
\includegraphics[height=3.18in,angle=-90]{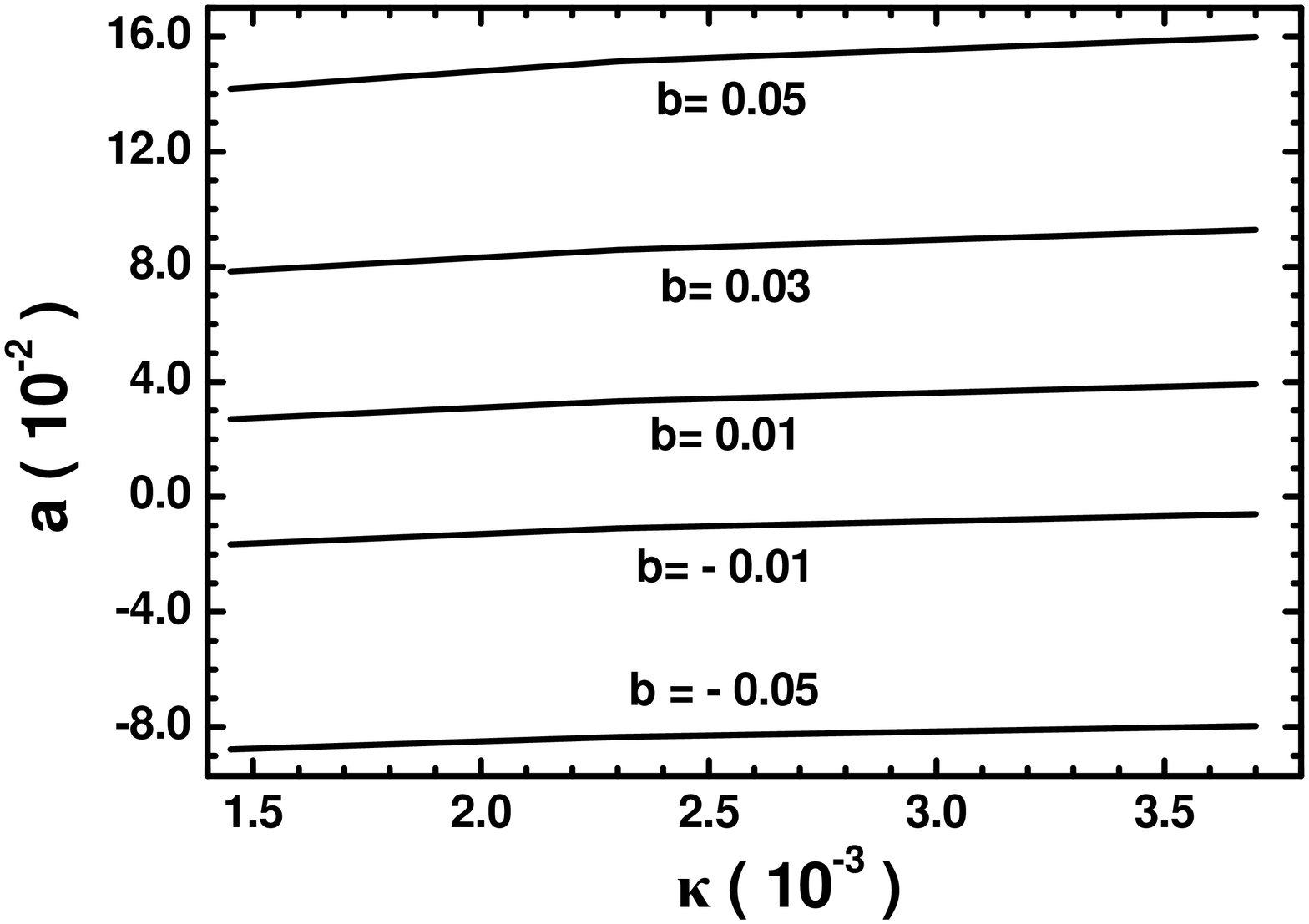}\hfil
\includegraphics[height=3.18in,angle=-90]{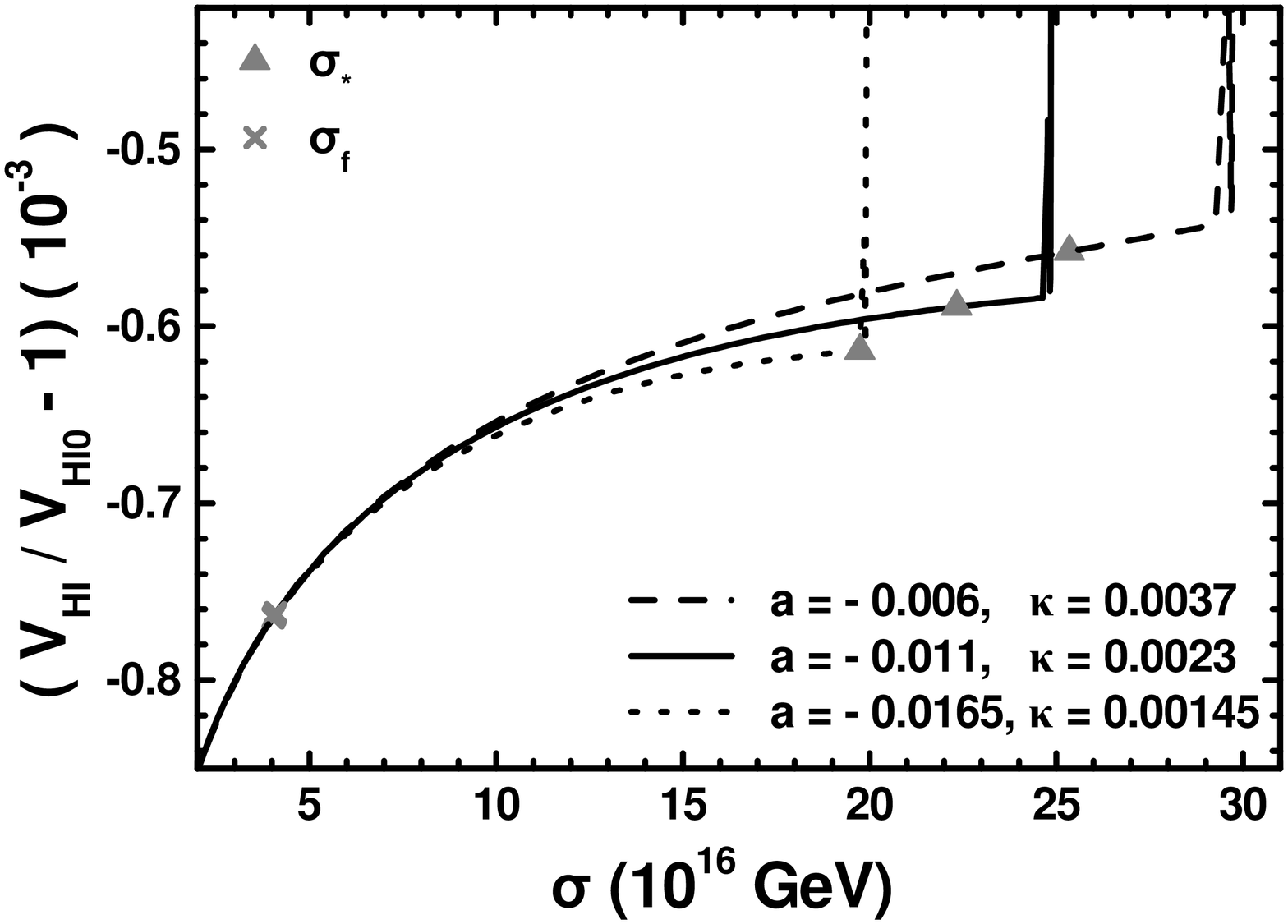}
\caption{\sl Allowed values by the requirements of Sec. 5.1 in the
$\kappa-\ck$ plane for various $b$'s indicated on the curves
(left) and the variation of $\Vhi$ as a function of $\sg$ for
$b=-0.01$ and $(\ck,\kp)$'s indicated in the graph (right). In
both graphs we use the values of the parameters shown in
Eq.~(5.4). For the right graph we set $f_{\rm
HIi}=3\cdot10^{17}~\GeV$ ($\ns=0.992$, dashed line) or $f_{\rm
HIi}=2.5\cdot10^{17}~\GeV$ ($\ns=0.968$, solid line) or $f_{\rm
HIi}=2\cdot10^{17}~\GeV$ ($\ns=0.944$, dotted line). The values
corresponding to $\sigma_*$ and $\sigma_{\rm f}$ are also
depicted.}\label{fig4}
\end{figure}

Letting $\ck$ vary for a number of fixed values of $b$, we can
depict the values allowed  by all the constraints of
Sec.~\ref{cont} in the $\kp-\ck$ plane -- see the left plot of
Fig.~\ref{fig4}. The various lines terminate at low [high]
$\kappa$'s due to the saturation of Eq.~(\ref{nsw}) from below
[above]. We readily conclude that the allowed $(\ck,b)$'s for
fixed $\ns$ are almost $\kp$-independent. This is because $\msm$
is fixed too. In particular, for $\ns=0.944,~0.968$ and $0.992$,
we have $-\msm\simeq0.0179,~0.0125$ and $0.0078$ and $\kp=0.00125,
0.002$ and $0.0037$,  respectively. In all cases,
$\msp\simeq1.82$, $Y_{1\Gr}\simeq5\cdot10^{-9}$ and
$Y_{2\Gr}\simeq2\cdot10^{-14}$. Therefore, our scenario can be
realized for both signs of $\ck$ and $b$, contrary to the cases
studied in Ref.~\cite{gpp} where negative $b$'s are necessitated.
Also, compared the extracted $Y_{2\Gr}$'s with the bounds of
\Eref{Ygr}, we infer that $\Gr$ with masses even lower than
$10~{\rm TeV}$ become observationally safe.

One of the outstanding features of our proposal is that the
reduction of $\ns$ can be attained without disturbing the
monotonicity of the potential -- cf. Ref.~\cite{gpp}. This fact is
highlighted in the right plot of Fig.~\ref{fig4}, where we present
the variation of the inflationary potential $\Vhi$ as a function
of $\sg$, for $b=-0.01$ and three pairs of $\ck$ and $\kp$'s,
shown in the graph, corresponding to $\ns=0.944$ (dotted line),
$0.968$ (solid line) and $0.992$ (dashed line). The values
corresponding to $\sigma_*$ and $\sigma_{\rm f}$ are also
designed. We observe that for large $\sg$'s, $\Vhi$ develops an
oscillatory behavior due to the initial oscillations of $s$ and
$q$ -- see Fig.~\ref{fig1}. However, $\Vhi$ for lower $\sg$'s
remains monotonic and, therefore, no complications arise in the
realization of FHI.

\section{Conclusions}\label{con}

We showed that, combining FHI with a PQPT based on renormalizable
superpotential terms, we can obtain: {\sf \ftn (i)}
Observationally viable FHI at the SUSY GUT scale with natural
values, $\pm(0.01-0.1)$, for the model parameters; {\sf \ftn (ii)}
a simultaneous resolution of the strong CP and $\mu$ problems of
MSSM; {\sf \ftn (iii)} a second stage of reheating after PQPT,
which leads to observationally safe values of the $\Gr$ abundance.
An important prerequisite for all these is that the field, which
triggers PQPT, remains after FHI well above the PQ scale thanks to
{\sf \ftn (i)} its participation in the SUGRA and logarithmic
corrections during FHI and {\sf \ftn (ii)} the high reheat
temperature after the same period. A noteworthy open issue of our
scenario is this of baryogenesis which cannot be processed via
non-thermal leptogenesis \cite{leptohybrid} since the produced
lepton asymmetry after FHI is efficiently diluted.


\def\ijmp#1#2#3{{\emph{Int. Jour. Mod. Phys.}}
{\bf #1},~#3~(#2)}
\def\plb#1#2#3{{\emph{Phys. Lett.  B }}{\bf #1},~#3~(#2)}
\def\zpc#1#2#3{{Z. Phys. C }{\bf #1},~#3~(#2)}
\def\prl#1#2#3{{\emph{Phys. Rev. Lett.} }
{\bf #1},~#3~(#2)}
\def\rmp#1#2#3{{Rev. Mod. Phys.}
{\bf #1},~#3~(#2)}
\def\prep#1#2#3{\emph{Phys. Rep. }{\bf #1},~#3~(#2)}
\def\prd#1#2#3{{\emph{Phys. Rev. } D }{\bf #1},~#3~(#2)}
\def\npb#1#2#3{{\emph{Nucl. Phys.} }{\bf B#1},~#3~(#2)}
\def\npps#1#2#3{{Nucl. Phys. B (Proc. Sup.)}
{\bf #1},~#3~(#2)}
\def\mpl#1#2#3{{Mod. Phys. Lett.}
{\bf #1},~#3~(#2)}
\def\arnps#1#2#3{{Annu. Rev. Nucl. Part. Sci.}
{\bf #1},~#3~(#2)}
\def\sjnp#1#2#3{{Sov. J. Nucl. Phys.}
{\bf #1},~#3~(#2)}
\def\jetp#1#2#3{{JETP Lett. }{\bf #1},~#3~(#2)}
\def\app#1#2#3{{Acta Phys. Polon.}
{\bf #1},~#3~(#2)}
\def\rnc#1#2#3{{Riv. Nuovo Cim.}
{\bf #1},~#3~(#2)}
\def\ap#1#2#3{{Ann. Phys. }{\bf #1},~#3~(#2)}
\def\ptp#1#2#3{{Prog. Theor. Phys.}
{\bf #1},~#3~(#2)}
\def\apjl#1#2#3{{Astrophys. J. Lett.}
{\bf #1},~#3~(#2)}
\def\n#1#2#3{{Nature }{\bf #1},~#3~(#2)}
\def\apj#1#2#3{{Astrophys. J.}
{\bf #1},~#3~(#2)}
\def\anj#1#2#3{{Astron. J. }{\bf #1},~#3~(#2)}
\def\mnras#1#2#3{{MNRAS }{\bf #1},~#3~(#2)}
\def\grg#1#2#3{{Gen. Rel. Grav.}
{\bf #1},~#3~(#2)}
\def\s#1#2#3{{Science }{\bf #1},~#3~(#2)}
\def\baas#1#2#3{{Bull. Am. Astron. Soc.}
{\bf #1},~#3~(#2)}
\def\ibid#1#2#3{{\it ibid. }{\bf #1},~#3~(#2)}
\def\cpc#1#2#3{{Comput. Phys. Commun.}
{\bf #1},~#3~(#2)}
\def\astp#1#2#3{{Astropart. Phys.}
{\bf #1},~#3~(#2)}
\def\epjc#1#2#3{{Eur. Phys. J. C}
{\bf #1},~#3~(#2)}
\def\nima#1#2#3{{Nucl. Instrum. Meth. A}
{\bf #1},~#3~(#2)}
\def\jhep#1#2#3{{\emph{JHEP} }
{\bf #1},~#3~(#2)}
\def\jcap#1#2#3{{\emph{JCAP} }
{\bf #1},~#3~(#2)}

\end{document}